\documentclass[aps, pra, twocolumn, amsmath, amssymb, superscriptaddress, nofootinbib]{revtex4-1}

\makeatletter
\def\p@subsection{}
\def\p@subsubsection{}
\makeatother

\usepackage[utf8]{inputenc}
\usepackage{graphicx}
\usepackage{units}
\usepackage{color}
\usepackage[pdftex, colorlinks=true, linkcolor=myblue, citecolor=myblue,
urlcolor=myblue]{hyperref}
\usepackage{times}
\usepackage{comment}
\usepackage{graphicx}
\usepackage{amsmath, calc}
\usepackage{mathrsfs}
\usepackage{amsfonts}
\usepackage{amssymb}
\usepackage{bm}
\usepackage{bbm}
\usepackage{color}
\usepackage{array}
\usepackage{units}
\usepackage{tabstackengine}
\stackMath

\definecolor{myblue}{rgb}{0,0,1}
\definecolor{myred}{rgb}{1,0,0}

\newcommand{\ket}[1]{|#1\rangle}

\DeclareMathOperator{\Tr}{Tr}

\newcommand\scalemath[2]{\scalebox{#1}{\mbox{\ensuremath{\displaystyle #2}}}}

\begin{document}
%\narrowtext

%===========================================================================
%===========================================================================
%===========================================================================
%===========================================================================
%===========================================================================
%===========================================================================
%===========================================================================
%===========================================================================

\title{Unbalanced gain and loss in a quantum photonic system}

%===========================================================================
%===========================================================================
%===========================================================================
%===========================================================================
%===========================================================================
%===========================================================================
%===========================================================================
%===========================================================================

\author{Charles Andrew Downing}
\email{c.a.downing@exeter.ac.uk} 
\affiliation{Department of Physics and Astronomy, University of Exeter, Exeter EX4 4QL, United Kingdom}

\author{Oliver Isaac Reuben Fox}
\affiliation{Department of Physics and Astronomy, University of Exeter, Exeter EX4 4QL, United Kingdom}

%===========================================================================
%===========================================================================
%===========================================================================
%===========================================================================
%===========================================================================
%===========================================================================
%===========================================================================
%===========================================================================

\date{\today}

%===========================================================================
%===========================================================================
%===========================================================================
%===========================================================================
%===========================================================================
%===========================================================================
%===========================================================================
%===========================================================================

\begin{abstract}
\noindent \textbf{Abstract}\\
Theories in physics can provide a kind of map of the physical system under investigation, showing all of the possible types of behavior which may occur. Certain points on the map are of greater significance than others, because they describe how the system responds in a useful or interesting manner. For example, the point of resonance is of particular importance when timing the pushes onto a person sat on a swing. More sophisticatedly, so-called exceptional points have been shown to be significant in optical systems harbouring both gain and loss, as typically described by non-Hermitian Hamiltonians. However, expressly quantum points of interest -- be they exceptional points or otherwise -- arising in quantum photonic systems have been far less studied. Here we consider a paradigmatic model: a pair of coupled qubits subjected to an unbalanced ratio of gain and loss. We mark on its map several flavours of both exceptional and critical points, each of which are associated with unconventional physical responses. In particular, we uncover the points responsible for characteristic spectral features and for the sudden loss of quantum entanglement in the steady state. Our results provide perspectives for characterizing quantum photonic systems beyond effective non-Hermitian Hamiltonians, and suggest a hierarchy of intrinsically quantum points of interest.
\\
\\ Keywords: non-Hermitian Hamiltonian, exceptional points, quantum master equation, qubits, open quantum systems.
\end{abstract}

%===========================================================================
%===========================================================================
%===========================================================================
%===========================================================================
%===========================================================================
%===========================================================================
%===========================================================================
%===========================================================================

\maketitle

%\pacs{73.22.Pr, 73.21.La, 03.65.Ge, 03.65.Pm}

%===========================================================================
%===========================================================================
%===========================================================================
%===========================================================================
%===========================================================================
%===========================================================================
%===========================================================================
%===========================================================================

\noindent \textbf{Introduction}\\
Conventionally, quantum mechanics deals with with Hermitian Hamiltonians. This common restriction ensures both real eigenvalues and unitary time evolution. Remarkably, relaxing this constraint by employing non-Hermitian Hamiltonians can also lead to a physical quantum theory~\cite{Bender2007, Moiseyev2011, Ashida2020}. Most famously, constructing a quantum theory obeying the less strict condition of parity-time (or $\mathcal{PT}$) symmetry has led to some spectacular theoretical predictions within modern optics~\cite{Konotop2016, Ganainy2018, Miri2019, Ozdemir2019}. In particular, theoretical treatments of gain and loss naturally lead to non-Hermitian Hamiltonians, which can become defective at so-called exceptional points -- points in parameter space which are typically associated with unconventional physics~\cite{Berry2003, Heiss2003}. The creation of popular non-Hermitian theories were soon followed by groundbreaking experimental successes in (mostly classical) optical systems demonstrating exceptional point physics~\cite{Ruter2010, Feng2014, Hodaei2014, Hodaei2017, Wu2019, Xia2021, Sweeney2021, Ergoktas2022}.

\begin{figure*}[tb]
 \includegraphics[width=1.0\linewidth]{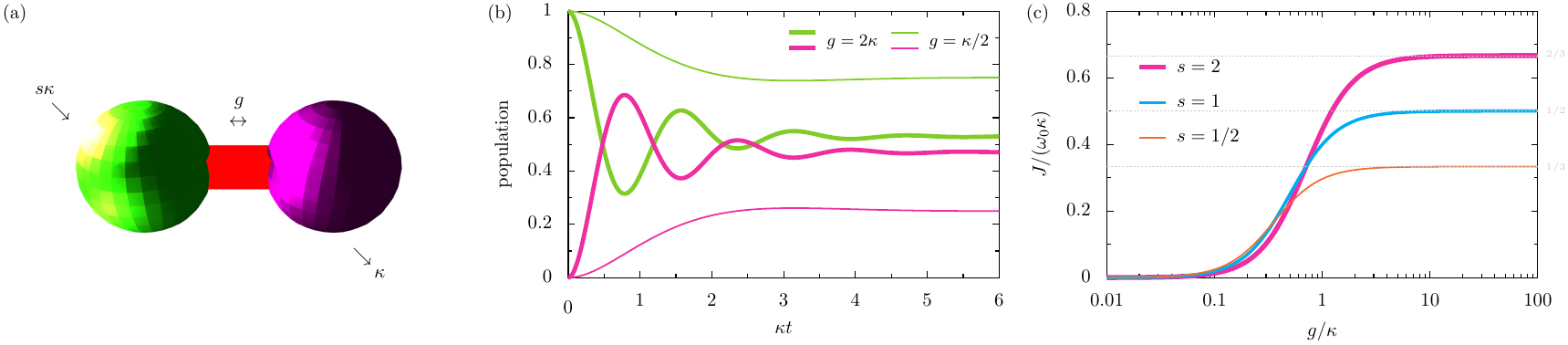}
 \caption{ \textbf{The unbalanced dimer and its populations.} Panel (a): a pair of two-level systems, each of transition frequency $\omega_0$, where the first qubit is pumped at a rate $s \kappa$, and the second qubit suffers loss at a rate $\kappa$. The coupling strength is $g$, and when the dimensionless parameter $s=1$ the system is $\mathcal{PT}$-symmetric. Panel (b): the mean populations in the first (green) and second (pink) qubits as a function of time $t$ (in units of $\kappa^{-1}$), for the cases of $g=2\kappa$ (thick lines) and $g=\kappa/2$ (thin lines), with $s=1$ [cf. Eq.~\eqref{eq:eigenfrsdfsdfequencies}]. Panel (c): the steady state current $J$ as a function of the coupling strength $g$ (in units of $\omega_0 \kappa$), for the cases of pumping associated with $s = \{ 1/2, 1, 2 \}$ [cf. Eq.~\eqref{eq:curr}]. }
 \label{popp}
\end{figure*}

Given the spectacular progress in $\mathcal{PT}$-symmetric optics, it is interesting to contemplate the consequences of non-Hermitian physics in essentially quantum optical systems~\cite{Prosen2012, Prosen2012b, Shallem2015, Kepesidis2016, Minganti2019, Jaramillo2020, Arkhipov2020, Downing2021}. Indeed, open quantum systems are a natural playground to study non-Hermitian effects due to their intrinsic description of system-environment interactions accounting for gain and loss~\cite{Huber2020, Khandelwal2021}. Perhaps the simplest case to consider is that of two coupled qubits~\cite{Huber2020, Khandelwal2021, Lopez2020, Supplementary}, as described by the Hamiltonian $\hat{H}$ (we take $\hbar = 1$)
\begin{equation}
\label{eq:Hammy}
 \hat{H} =~\omega_0 \left( \sigma_1^{\dagger} \sigma_1 +  \sigma_2^{\dagger} \sigma_2 \right) +  g \left( \sigma_1^{\dagger} \sigma_2 + \sigma_2^{\dagger} \sigma_1 \right),
\end{equation}
where $\omega_0$ is the common transition frequency of both two-level systems (2LSs), and the coupling strength between them is accounted with $g \ge 0$. The raising and lowering operators $\sigma_n^{\dagger}$ and $\sigma_n$ satisfy both the anticommutator relation $\{ \sigma_n, \sigma_n^\dagger \} = 1$ and the rule $\sigma_1 \sigma_2 = \sigma_2 \sigma_1$, corresponding to two distinguishable qubits~\cite{Allen1975, Zueco2021}. Accounting for losses in an open quantum systems approach, we may employ the following quantum master equation for the system's density matrix $\rho$~\cite{Gardiner2014,Supplementary}
\begin{align}
\label{eqapp:massdsdsdter}
 \partial_t \rho =&~~\mathrm{i} [ \rho, \hat{H} ] \nonumber \\
&+ \frac{s\kappa}{2} \left( 2 \sigma_1^{\dagger} \rho \sigma_1 -  \sigma_1 \sigma_1^{\dagger} \rho - \rho \sigma_1 \sigma_1^{\dagger} \right) \nonumber \\
&+ \frac{\kappa}{2} \left( 2 \sigma_2 \rho \sigma_2^{\dagger} -  \sigma_2^{\dagger} \sigma_2 \rho - \rho \sigma_2^{\dagger} \sigma_2 \right),  
\end{align}
where the rate $\kappa \ge 0$ measures the loss out of the second 2LS, and the dimensionless parameter $s \ge 0$ tunes the overall gain rate $s\kappa$ into the first 2LS. Hence the gain-loss ratio amongst the coupled qubits is generally unbalanced (that is, as long as $s \ne 1$) as is sketched schematically in Fig.~\ref{popp}~(a).

The time evolution of the density matrix $\rho$ in Eq.~\eqref{eqapp:massdsdsdter} is governed by three ingredients. On the right-hand side of Eq.~\eqref{eqapp:massdsdsdter}, the first line contains the Liouville–von Neumann commutator, involving the Hermitian Hamiltonian operator $\hat{H}$ of Eq.~\eqref{eq:Hammy}. The second line in Eq.~\eqref{eqapp:massdsdsdter} alludes to incoherent gain into the first qubit, and the third line in Eq.~\eqref{eqapp:massdsdsdter} contains the Lindbladian dissipator of the second qubit -- these two additions impart non-Hermiticity into the model. Notably, we do not remove the so-called refilling (or feeding) terms from the two Lindbladians featuring in Eq.~\eqref{eqapp:massdsdsdter}, and so we do not entertain the celebrated case of a traditional non-Hermitian Hamiltonian model~\cite{Supplementary}. Nevertheless, the quantum physics of non-Hermitian Hamiltonians is rather interesting, including for the promotion of squeezing and for the enhancement of quantum entanglement~\cite{Reiter2014, Ramirez2019}.

We also note that the quantum optical master equation formalism that we employ in Eq.~\eqref{eqapp:massdsdsdter} is essentially equivalent to a more technically demanding quantum Langevin equation-style approach, since both approaches are unifiable within a higher-level quantum noise theory~\cite{Gardiner1988}.

This considered minimal model of two coupled qubits, as encapsulated by Eq.~\eqref{eq:Hammy} and Eq.~\eqref{eqapp:massdsdsdter}, showcases a variety of quantum points of interest -- including both exceptional points and critical points. These special points arise when the system parameters approach the values
\begin{subequations}
\label{eq:xvcxvcxvc}
\begin{align}
 s &= 1, \quad &&\mathcal{PT}~\mathrm{symmetry}, \label{eq:xvcxvcxvcA} \\
    g &= \sqrt{s} \left( \frac{\sqrt{2}\pm1}{2} \right) \kappa, &&\mathrm{type-I~MEP}, \label{eq:xvcxvcxvcB} \\
  g &= \frac{|s-1|}{4} \kappa, &&\mathrm{type-II~MEP}, \label{eq:xvcxvcxvcC} \\
   g &= \left( \frac{s+1}{2} \right) \kappa, &&\mathrm{SCP}. \label{eq:xvcxvcxvcD}
  \end{align}
\end{subequations}
Let us consider each quantum point of interest in the list of Eq.~\eqref{eq:xvcxvcxvc} in turn:

\emph{A}. In essentially classical systems, the border case between a system being open and being closed has received considerable attention. It was shown by Bender and co-workers~\cite{Bender1998, Bender2018} that when $\mathcal{PT}$-symmetry is obeyed by an effective Hamiltonian, the resulting eigenvalues can be real despite the Hamiltonian being non-Hermitian. The prototypical $\mathcal{PT}$-symmetric system is that of two coupled harmonic oscillators~\cite{Ozdemir2019}, one being supplied with gain $\kappa$, and the other suffering loss at the equivalent rate $\kappa$. Then it follows that upon carrying out the parity operation (switching the positions of the two oscillators) and the time operation (flipping time, so that gain and loss are interchanged) the transformed system is equivalent to the starting one: thus $\mathcal{PT}$-symmetry is obeyed~\cite{Ozdemir2019}. It was recently proposed by Huber and co-workers~\cite{Huber2020} that the natural operation extending $\mathcal{PT}$-symmetry to open quantum systems is to interchange loss and gain using the operator transformations $\sigma_1 \to \sigma_1^\dagger$ and $\sigma_2 \to \sigma_2^\dagger$ within the Lindblad dissipators. Hence, for the setup we consider in Fig.~\ref{popp}~(a), the system exhibits $\mathcal{PT}$ symmetry when the dimensionless parameter $s=1$, such that the gain and loss is indeed evenly balanced [cf. Eq.~\eqref{eq:xvcxvcxvcA} and Fig.~\ref{popp}~(a)].
  
\emph{B}.  An intuitive way to write down an effective Hamiltonian $\mathcal{H}$ arising from a quantum master equation is to form the matrix equation for the mean values of the first moments, that is objects like $\langle \sigma_1 \rangle$ and $\langle \sigma_2 \rangle$. In the case of the considered Eq.~\eqref{eqapp:massdsdsdter}, the resultant effective dynamical equation is~\cite{Supplementary}
  \begin{equation}
\label{eqapp:oasdadsf_motion_io}
\mathrm{i} \partial_t \psi = \mathcal{H} \psi,
\end{equation}
 where the effective Hamiltonian $\mathcal{H}$ is a $4\times4$ matrix
 \begin{equation}
\label{eqapp:oasfdsdadf_motion_iwewo}
\fixTABwidth{T}
\scalemath{0.855}{
 \mathcal{H} =
\begin{pmatrix}
\omega_{0} - \mathrm{i} \frac{s \kappa}{2} & g & -2g & 0 \\ 
g  & \omega_{0} - \mathrm{i} \frac{\kappa}{2} & 0 & -2g\\ 
0 & \mathrm{i} s \kappa & \omega_{0}  - \mathrm{i} \kappa \left( s + \frac{1}{2} \right) & -g \\ 
0 & 0 & -g & \omega_{0}  - \mathrm{i} \kappa \left( 1 + \frac{s}{2} \right)
 \end{pmatrix}},
\end{equation}
and where $\psi = ( \langle \sigma_1 \rangle, \langle \sigma_2 \rangle, \langle \sigma_1^{\dagger} \sigma_1 \sigma_2 \rangle, \langle \sigma_1 \sigma_2^{\dagger} \sigma_2  \rangle )^\mathrm{T}$ collects the averaged first moment operators. The four complex eigenvalues of $\mathcal{H}$ (and their associated eigenvectors) coalesce at two distinct points, exactly when the conditions of Eq.~\eqref{eq:xvcxvcxvcB} are met --  they are truly exceptional points or $\mathrm{EPs}$. We call these $\mathrm{EPs}$ arising from the first moments of the system $\mathrm{type-I~MEP}$s. Strikingly, these $\mathrm{EPs}$ exist both in the $\mathcal{PT}$-symmetric arrangement ($s = 1$) and outside this balanced setup ($s \ne 1$). 
 
\emph{C}.  Naturally, another appealing route to uncover $\mathrm{EPs}$ within an open quantum systems approach is to consider higher orders of the moments. For example, by considering the second moments of the system, objects like $\langle \sigma_1^\dagger \sigma_1 \rangle$ and $\langle \sigma_2^\dagger \sigma_2 \rangle$, the dynamical equation formed from the quantum master equation of Eq.~\eqref{eqapp:massdsdsdter} is given by
\begin{equation}
\label{eqapp:ssdsfd}
\mathrm{i }\partial_t  \Psi  = \mathcal{M} \Psi + \mathcal{P},
\end{equation}
where $\Psi = ( \langle \sigma_1^{\dagger} \sigma_1 \rangle, \langle \sigma_2^{\dagger}  \sigma_2 \rangle, \langle \sigma_1^{\dagger} \sigma_2 \rangle, \langle  \sigma_2^{\dagger} \sigma_1  \rangle )^\mathrm{T}$ gathers the four second moments, and with the column vector driving term $\mathcal{P} = ( \mathrm{i} s \kappa, 0, 0, 0 )^\mathrm{T}$. Explicitly, the dynamical matrix $ \mathcal{M}$ is defined via
 \begin{equation}
\label{eqapp:oasfsdfdsdadf_motsdfdsfion_iwewo}
 \mathcal{M} =
\begin{pmatrix}
  -\mathrm{i } s \kappa &  0 &  g & -  g  \\
    0 &   -\mathrm{i } \kappa &  -  g &  g  \\
       g &  -  g &  -\mathrm{i } \left( \frac{1+s}{2} \right) \kappa & 0  \\
        -  g &    g &  0 & -\mathrm{i } \left( \frac{1+s}{2} \right) \kappa
 \end{pmatrix},
\end{equation}
which admits complex eigenvalues (and eigenvectors) which simultaneously coalescence at only one point [cf. Eq.~\eqref{eq:xvcxvcxvcC}]. This $\mathrm{EP}$ may be termed a $\mathrm{type-II~MEP}$ due to its association with the second moments. Notably, here the $\mathrm{EP}$ vanishes when $s=1$, such that there is no $\mathrm{type-II~MEP}$ when the system is in its $\mathcal{PT}$-symmetric, balanced arrangement. 
  
\emph{D}. The quantum nature of the model of Eq.~\eqref{eqapp:massdsdsdter} naturally leads one to consider the explicitly quantum features which may be governed by truly quantum points of interest. Considerations of the properties of the density matrix $\rho$ -- and especially entanglement measures -- lead to the observation of a critical (and non-exceptional) point. In particular, the critical point at which the quantum state suddenly becomes unentangled (or separable), which we dub a separable critical point or $\mathrm{SCP}$, is given by Eq.~\eqref{eq:xvcxvcxvcD} in the steady state.
 
Notably, if we were to neglect the refilling terms in the Lindbladians appearing in Eq.~\eqref{eqapp:massdsdsdter} in order to enter a non-Hermitian Hamiltonian model regime, the two influential dynamical matrices discussed previously, $\mathcal{H}$ and $\mathcal{M}$ as defined in Eq.~\eqref{eqapp:oasfdsdadf_motion_iwewo} and Eq.~\eqref{eqapp:oasfsdfdsdadf_motsdfdsfion_iwewo}] respectively, would be markedly different~\cite{Supplementary}. As a knock-on effect, the locations of the $\mathrm{EPs}$ (as well as the overall dynamics) also change significantly. However, there is the pronounced cost of the approximation employed (neglecting the refilling terms) highly restricting the parameters one may reasonably consider, such that we relegate the non-Hermitian Hamiltonian aspects of the model to the Supplementary Material~\cite{Supplementary}.  

In what follows, we consider the impact of the four quantum points of interest listed in Eq.~\eqref{eq:xvcxvcxvc} in a handful of common observables and measures. In doing so, we highlight the importance of such analyses for complete descriptions of quantum optical systems with substantial non-Hermitian aspects (as they arise from a proper quantum master equation approach).
\\

\begin{figure*}[tb]
 \includegraphics[width=1.0\linewidth]{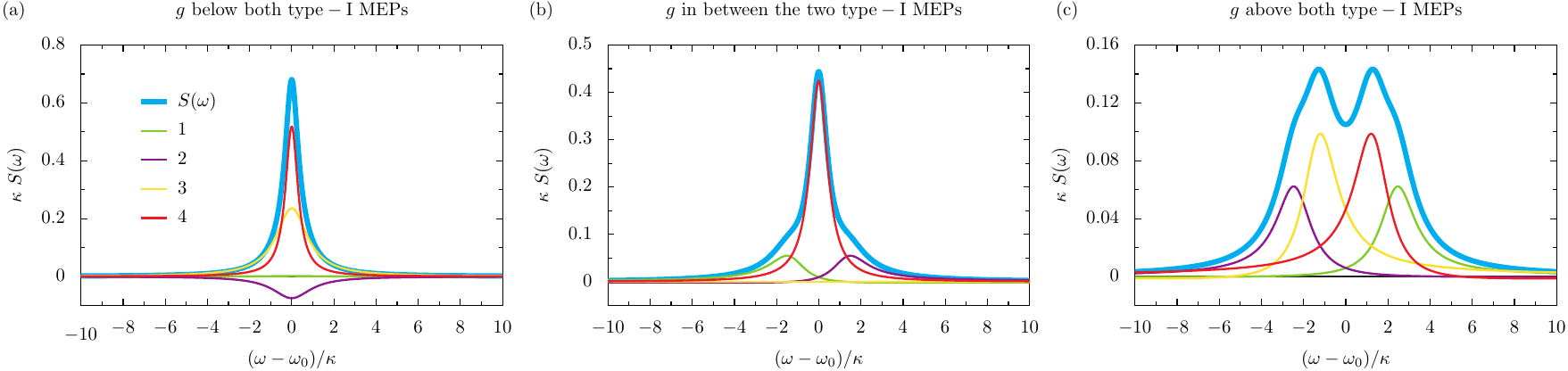}
 \caption{ \textbf{The unbalanced dimer and its spectrum.} The optical spectrum $S(\omega)$, as a function of the shifted emission frequency $\omega-\omega_0$ (in units of $\kappa$), for the balanced case of $s = 1$ (thick cyan lines). Panel (a): the coupling strength $g = \kappa/5$, which is below the first $\mathrm{type-I~MEP}$. Panel (b): $g = \kappa$, which is between the first and second $\mathrm{type-I~MEP}$s. Panel (c): $g = 2\kappa$, which is above the second and final $\mathrm{type-I~MEP}$. Thin colored lines: the four contributions to the overall spectrum, due to the four possible transitions in the system [cf. Eq.~\eqref{eqapp:sasdsdfsdfsdfccasd}]. The $\mathrm{type-I~MEP}$s occur at $g = (\sqrt{2}-1)\kappa/2 \simeq 0.207\kappa$ and $g = (\sqrt{2}+1)\kappa/2 \simeq 1.207\kappa$ respectively [cf. Eq.~\eqref{eq:xvcxvcxvcB}].}
 \label{nelson}
\end{figure*}

%===========================================================================
%===========================================================================
%===========================================================================
%===========================================================================
%===========================================================================
%===========================================================================
%===========================================================================
%===========================================================================

\noindent \textbf{Populations}\\
The mean values of the second moments $\langle \sigma_n^\dagger\sigma_n\rangle$ gives access to the average populations of the two qubits. The analytic expressions for the dynamic populations, with the initial conditions $\langle \sigma_1^\dagger\sigma_1\rangle = 1$ and $\langle \sigma_2^\dagger\sigma_2\rangle = 0$ at the starting time $t=0$, are particularly simple when the loss and gain are balanced (that is, when $s=1$). These two qubit populations, accessed from the solution of the equation of motion given in Eq.~\eqref{eqapp:ssdsfd}, then read~\cite{Supplementary, Sturges2022}
\begin{subequations}
\label{eq:eigenfrsdfsdfequencies}
\begin{align}
 \langle \sigma_1^\dagger\sigma_1\rangle &= \frac{\kappa^2+2g^2}{\kappa^2+4g^2} + \frac{  2 g^2 \cos \left( 2 g t \right) + \kappa g \sin \left( 2 g t \right)}{\kappa^2+4g^2}  \mathrm{e}^{-\kappa t}, \label{eq:eigenfsdrequenciesC}\\
 \langle \sigma_2^\dagger\sigma_2\rangle &= \frac{2g^2}{\kappa^2+4g^2} - \frac{  2g^2 \cos \left( 2 g t \right) + \kappa g \sin \left( 2 g t \right)}{\kappa^2+4g^2} \mathrm{e}^{-\kappa t},\label{eq:eigenfrequsdenciesB} 
  \end{align}
\end{subequations}
which display exponential decay, with the time constant $1/\kappa$, of the trigonometric population oscillations towards unequal steady states [given by the first terms on the right-hand sides of Eq.~\eqref{eq:eigenfrsdfsdfequencies}]. Such behaviour for the left (green) and right (pink) qubits is displayed graphically in Fig.~\ref{popp}~(b). When the coupling $g$ is strong (thick lines) distinctive Rabi cycles of the populations may be observed. Otherwise, for weak coupling $g$ (thin lines) there is a monotonic approach to the steady state population, which is created due to an equilibrium being found between the competing loss and gain processes.

The results for unbalanced arrangements of the qubits (where the dimensionless parameter $s \ne 1$) do indeed admit $\mathrm{type-II~MEP}$s [cf. Eq.~\eqref{eq:xvcxvcxvcC}]. However, the reconstruction of the population expressions of Eq.~\eqref{eq:eigenfrsdfsdfequencies} around these exceptional points (from damped-trigonometric to damped-algebraic) is not readily noticeable and so these results are not displayed here -- although we do note that some signatures of these $\mathrm{EP}$s have recently been obtained in Ref.~\cite{Khandelwal2021}.

The general analytic expressions for the steady state populations $\lim_{t \to \infty} \langle \sigma_n^\dagger\sigma_n\rangle$ are readily obtainable for any gain-loss imbalance, as judged by the dimensionless parameter $s$. The following neat formula for the population imbalance across the dimer system may then be found~\cite{Supplementary}
\begin{equation}
\label{eq:imb}
\lim_{t\to\infty} \frac{\langle \sigma_1^\dagger\sigma_1\rangle - \langle \sigma_2^\dagger\sigma_2\rangle}{ \langle \sigma_1^\dagger\sigma_1\rangle + \langle \sigma_2^\dagger\sigma_2\rangle }  = \frac{ \left( s+1\right) \kappa^2  }{ 8 g^2 + \left( s+1\right) \kappa^2  },
\end{equation}
which is bounded between $0$ and $1$ for strong and weak couplings $g$ respectively. The population imbalance of Eq.~\eqref{eq:imb} suggests the formation of a steady state current across the duo of qubits is possible~\cite{Manzano2012}. Indeed, the driving $s \kappa$ into the first 2LS and the loss $\kappa$ out of the second 2LS allow for a nonzero steady state current $J$ to be set up, with the strength~\cite{Supplementary}
\begin{equation}
\label{eq:curr}
J = \left( \frac{s}{s+1} \right) \frac{4 g^2 \kappa }{4 g^2 + s \kappa^2}~ \omega_0.
\end{equation}
The current $J$ displays the bounds $0 \le J \le \omega_0 \kappa s/(s+1)$ for weak and strong couplings $g$, as shown in Fig.~\ref{popp}~(c) for several values of $s$. This current $J$ could act as a useful observable to calibrate the coupled quantum system under experimental consideration.
\\

%===========================================================================
%===========================================================================
%===========================================================================
%===========================================================================
%===========================================================================
%===========================================================================
%===========================================================================
%===========================================================================

\noindent \textbf{Spectra}\\
The quantum regression formula provides a route to the first-order correlation function $g_n^{(1)} (\tau)$ of the $n$-th qubit. With the delay time $\tau \ge 0$, this degree of coherence may be defined, using the steady state population as normalization, as $g_n^{(1)} (\tau) = \lim_{t \to \infty} \langle \sigma_n^\dagger(t) \sigma_n (t+\tau) \rangle / \langle \sigma_n^\dagger(t) \sigma_n(t) \rangle$. By definition, $|g_n^{(1)} (0) | = 1$ demonstrates full coherence, $|g_n^{(1)} (0) | = 0$ perfect incoherence and otherwise $0 < |g_n^{(1)} (0) | < 1$ partial coherence~\cite{Gardiner2014}. Analytical formulas for $g_n^{(1)} (\tau)$ are rather long in general, but simplify in the balanced case of $s=1$ to~\cite{Supplementary}
\begin{widetext}
\begin{subequations}
\label{eq:dfdfdfdf}
\begin{align}
 g_1^{(1)} (\tau) &= \frac{\mathrm{e}^{- \mathrm{i} \omega_0 \tau} \mathrm{e}^{-\kappa \tau}}{\kappa^2+2g^2}
 \bigg\{ \frac{\kappa^2 + \kappa g + 2 g^2}{2}  \cos \left( q \tau \right) +  \frac{\kappa^2 - \kappa g + 2 g^2}{2} \cos \left( p \tau \right)
  +  \frac{\kappa^2 \left( 3 g + \kappa \right) }{4 q} \sin \left( q \tau \right) -  \frac{\kappa^2 \left( 3 g - \kappa \right) }{4 p} \sin \left( p \tau \right) \bigg\}, \label{eq:gkkh}\\
 g_2^{(1)} (\tau) &= \frac{\mathrm{e}^{- \mathrm{i} \omega_0 \tau} \mathrm{e}^{-\kappa \tau}}{4g}
 \bigg\{ \left( 2 g + \kappa \right) \cos \left( q \tau \right) +  \left( 2 g - \kappa \right) \cos \left( p \tau \right)
  +  \frac{\kappa \left( 4 g + \kappa \right) }{2 q} \sin \left( q \tau \right) +  \frac{\kappa \left( 4 g - \kappa \right) }{2 p} \sin \left( p \tau \right) \bigg\}, \label{eq:ng}
  \end{align}
\end{subequations}
\end{widetext}
where we have introduced the auxiliary functions $q = \sqrt{g^2 - g \kappa -\kappa^2/4}$ and $p = \sqrt{g^2 + g \kappa -\kappa^2/4}$. Total incoherence arises at long times, $g_n^{(1)} (\infty) = 0$, due to the lack of correlations with large time delays $\tau$.

The optical spectrum $S (\omega)$ of the first (and driven) qubit may be calculated directly from the degree of coherence $g_1^{(1)} (\tau)$, via the normalized spectral formula $S (\omega) = \lim_{t \to \infty} \mathrm{Re} \int_0^\infty g_1^{(1)} (\tau) / \pi$~\cite{Valle2010, Downing2023}. There are four possible transitions in the coupled system (in the simplest case of the bare states, they are from $\ket{1, 1} \to \ket{1, 0} \: \& \: \ket{0, 1}$ and from $\ket{1, 0} \: \& \: \ket{0, 1} \to \ket{0, 0}$) which are all associated with a certain transition frequency and a certain lifetime. Therefore, there are four contributions (thin colored lines) to the overall spectrum $S (\omega)$ (thick cyan lines) with particular peak positions and broadenings, as is shown in Fig.~\ref{nelson}~(a, b, c). The positions and broadenings may be determined via the real and imaginary parts respectively of the four complex eigenvalues $\epsilon_{\pm, \pm}$, arising from the governing dynamical matrix $\mathcal{H}$ of Eq.~\eqref{eqapp:oasfdsdadf_motion_iwewo}, which read~\cite{Supplementary}
\begin{equation}
\label{eqapp:sasdsdfsdfsdfccasd}
\epsilon_{\pm, \pm} = \omega_0 - \mathrm{i} \kappa \pm \sqrt{g^2 - \left( \frac{\kappa}{2} \right)^2 \pm g \kappa}.
\end{equation}
This expression contains two points of coalescence, consistent with the $\mathrm{type-I~MEP}$s of Eq.~\eqref{eq:xvcxvcxvcB}. Across the row of panels in the spectral Fig.~\ref{nelson}, we consider three successively increasing values of the coupling strength $g$ for the balanced case of $s=1$, so that the $\mathrm{type-I~MEP}$s occur at $g = (\sqrt{2}-1)\kappa/2 \simeq 0.207 \kappa$ and $g = (\sqrt{2}+1)\kappa/2 \simeq 1.207 \kappa$ [cf. Eq.~\eqref{eq:xvcxvcxvcB}]. In panel (a) the coupling strength $g$ is below the first $\mathrm{type-I~MEP}$, and the spectrum $S (\omega)$ presents a singlet structure since the complex eigenvalues describing the transitions all have $\omega_0$ as the real part and four distinct imaginary parts leading to four different spectral broadenings [cf. Eq.~\eqref{eqapp:sasdsdfsdfsdfccasd}]. Between the first $\mathrm{type-I~MEP}$ and the second $\mathrm{type-I~MEP}$, there are three distinct real parts of the complex eigenvalues (since the first $\mathrm{type-I~MEP}$ has been passed, associated with a coalescence) such that the singlet has developed noticeable spectral shoulders in panel (b). Finally in panel (c), above the second and final $\mathrm{type-I~MEP}$, the four distinct real parts of the complex eigenvalues [cf. Eq.~\eqref{eqapp:sasdsdfsdfsdfccasd}] lead to a characteristic doublet spectrum with significant spectral shoulders. Hence the evolution with $g$ of the number of peaks within the optical spectrum $S (\omega)$ presents a useful and rather vivid indicator for when the system has passed through $\mathrm{type-I~MEP}$s.
\\

%===========================================================================
%===========================================================================
%===========================================================================
%===========================================================================
%===========================================================================
%===========================================================================
%===========================================================================
%===========================================================================

\begin{figure*}[tb]
 \includegraphics[width=1.0\linewidth]{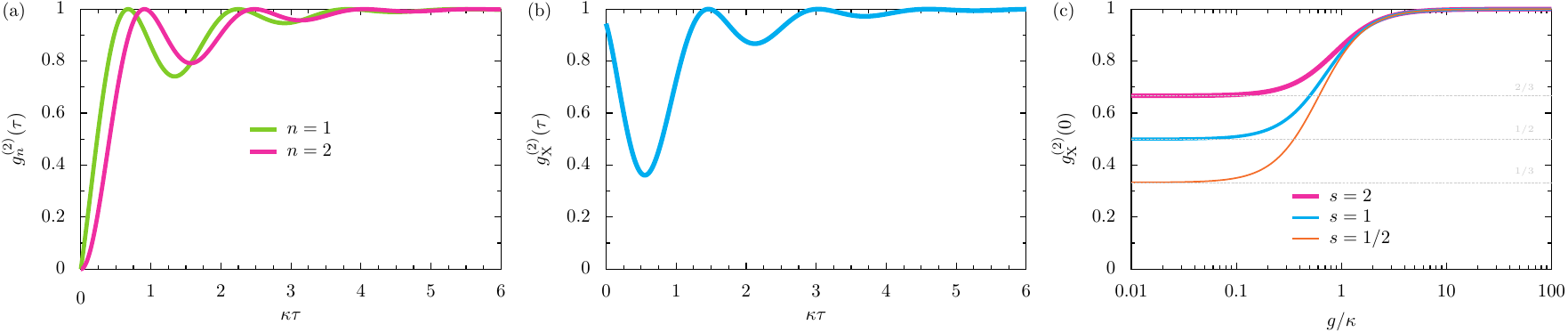}
 \caption{ \textbf{The unbalanced dimer and its correlations.} Panel (a): second-order direct correlation function $g_n^{(2)} (\tau)$ as a function of the delay time $\tau$ (in units of $\kappa^{-1}$) for the first ($n=1$) and second ($n=2$) resonators [cf. Eq.~\eqref{eq:eigenfsdrezxczxcquenciesA} and Eq.~\eqref{eq:eigenfsdrezxczxcquenciesB}], with $s=1$. Panel (b): second-order cross-correlation function $g_{\mathrm{X}}^{(2)} (\tau)$, as a function of the delay time $\tau$, with $s=1$ [cf. Eq.~\eqref{eq:rtrt}]. Panel (c): second-order cross-correlation function at zero delay $g_{\mathrm{X}}^{(2)} (0)$, as a function of the coupling strength $g$ (in units of $\kappa$), for the cases of $s = \{ 1/2, 1, 2 \}$ [cf. Eq.~\eqref{eq:rtrddft}]. }
 \label{corr}
\end{figure*}

\noindent \textbf{Correlations}\\
The degree of second-order coherences $g^{(2)} (\tau)$ measure the emission properties of the duo of coupled qubits~\cite{Gardiner2014}. The normalized second-order correlation functions are defined by, for the two direct correlation functions: $g_n^{(2)} (\tau) = \lim_{t \to \infty} \langle \sigma_n^\dagger(t) \sigma_n^\dagger(t+\tau) \sigma_n (t+\tau) \sigma_n (t) \rangle / \langle \sigma_n^\dagger(t) \sigma_n(t) \rangle^2$, with $n = \{ 1, 2\}$. These correlators count the probabilities of two emissions, with a time delay $\tau$, from the same qubit. Likewise, the cross-correlator is similarly given by $ g_{\mathrm{X}}^{(2)} (\tau) = \lim_{t \to \infty} \langle \sigma_1^\dagger(t) \sigma_2^\dagger(t+\tau) \sigma_2 (t+\tau) \sigma_1 (t) \rangle / (\langle \sigma_1^\dagger(t) \sigma_1(t) \rangle \langle \sigma_2^\dagger(t) \sigma_2(t) \rangle )$, which tracks two $\tau$-delayed emissions, one from each qubit in the dimer.

The explicit expressions for these types of $g^{(2)} (\tau)$ are most simple in the balanced case of $s=1$, where they read~\cite{Supplementary}
\begin{subequations}
\label{eq:eigenfrsdfdsadsdsdfequencies}
\begin{align}
 g_1^{(2)} (\tau) &= 1 - \frac{\left[ \left( 2g^2+\kappa^2 \right) \cos \left(g \tau \right) - g \kappa \sin \left( g \tau \right) \right]^2}{\left( 2g^2+\kappa^2 \right)^2}  \mathrm{e}^{-\kappa \tau}, \label{eq:eigenfsdrezxczxcquenciesA}\\
 g_2^{(2)} (\tau) &= 1 - \frac{\left[ 2g \cos \left(g \tau \right) + \kappa \sin \left( g \tau \right) \right]^2}{4g^2}  \mathrm{e}^{-\kappa \tau}, \label{eq:eigenfsdrezxczxcquenciesB}\\
  g_{\mathrm{X}}^{(2)} (\tau) &= 1 - \frac{\left[ g\kappa \cos \left(g \tau \right) + \left( 2g^2+\kappa^2 \right) \sin \left( g \tau \right) \right]^2}{2g^2\left( 2g^2+\kappa^2 \right)}  \mathrm{e}^{-\kappa \tau}. \label{eq:rtrt}
  \end{align}
\end{subequations}
We plot the two direct correlation functions $g_n^{(2)} (\tau)$ in Fig.~\ref{corr}~(a), where $n=1$ is marked with the green line and $n=2$ by the pink line. At zero time delay ($\tau = 0$), the two-level nature of the qubits enforces $g_n^{(2)} (0) = 0$, since a finite time $\tau$ is required for the qubit to be re-excited from the ground state in order to emit a second photon. For nonzero $\tau$ the phenomena of antibunching, as defined by $g_n^{(2)} (0) < g_n^{(2)} (\tau)$ emerges as a non-classical effect. The long time limit ($\tau \to \infty$) sees perfect coherence being approached due to the random nature of the emissions in this asymptotic scenario.

We plot the cross-correlation function $g_{\mathrm{X}}^{(2)} (\tau)$, as defined in Eq.~\eqref{eq:rtrt}, in Fig.~\ref{corr}~(b). The result $g_{\mathrm{X}}^{(2)} (0) \ne 0$ arises since simultaneous emissions, one in each qubit, are possible even at zero time delay ($\tau = 0$). Thereafter, there are characteristic oscillations until the asymptotic result $g_{\mathrm{X}}^{(2)} (\infty) = 1$ is approached, signifying the qubits are behaving independently for large delay times $\tau$. At zero delay ($\tau = 0$), and for a general unbalanced system with any $s$, we find the cross-correlation formula
\begin{equation}
\label{eq:rtrddft}
   g_{\mathrm{X}}^{(2)} (0) = \frac{4 g^2 + s\kappa^2}{4g^2 + (s+1)\kappa^2}.
\end{equation}
We plot $g_{\mathrm{X}}^{(2)} (0)$ in Fig.~\ref{corr}~(c) as a function of the coupling $g$ for three values of $s$, showcasing the weak coupling limit of $g_{\mathrm{X}}^{(2)} (0) = 1/(1+s)$, and the strong coupling result of $g_{\mathrm{X}}^{(2)} (0) = 1$, such that the degree of antibunching can be somewhat tuned by the pumping strength $s \kappa$. Similar to the population dynamics, the correlation functions considered (with $s \ne 1$) do not allow for as striking a fingerprint of an exceptional point as compared to the optical spectrum, since the behaviour of the relevant functions at the exceptional point is rather similar to the behaviour immediately above and below it.
\\

%===========================================================================
%===========================================================================
%===========================================================================
%===========================================================================
%===========================================================================
%===========================================================================
%===========================================================================
%===========================================================================

\begin{figure*}[tb]
 \includegraphics[width=1.0\linewidth]{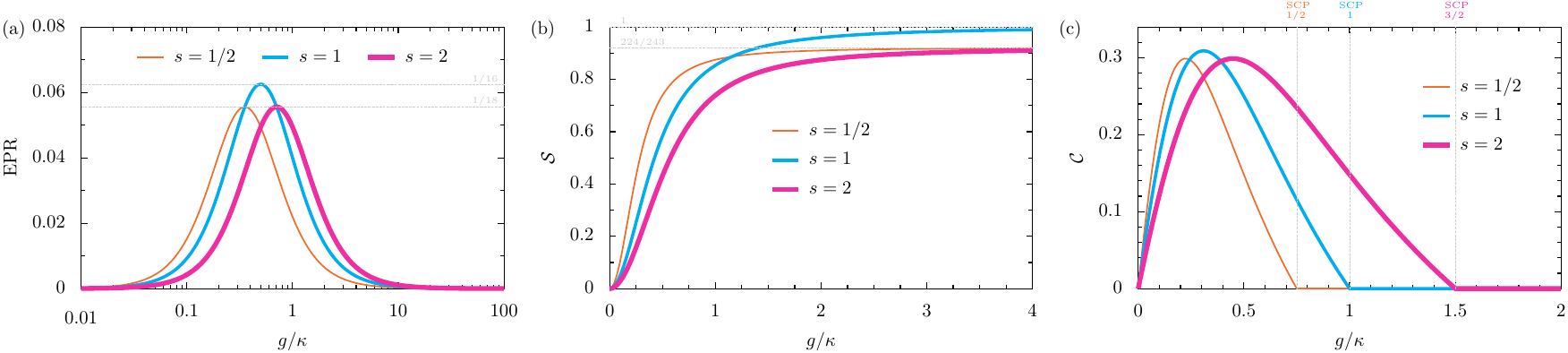}
 \caption{ \textbf{The unbalanced dimer and its quantum information in the steady state.} Panel (a): the entanglement measure $\mathrm{EPR}$ as a function of the coupling strength $g$ (in units of $\kappa$) for the cases of $s = \{ 1/2, 1, 2 \}$ [cf. Eq.~\eqref{eq:frt}]. Dashed grey lines: the maximum values of $1/16 \simeq 0.063$ and $1/18 \simeq 0.056$. Panel (b): the linear entropy $\mathcal{S}$ as a function of the scaled coupling strength $g/\kappa$ [cf. Eq.~\eqref{eq:lin}]. Dashed grey lines: the asymptotic values of $1$ and $224/243 \simeq 0.92$. Panel (c): the concurrence $\mathcal{C}$ as a function of the scaled coupling strength $g/\kappa$. Dashed grey lines: the concurrence zeroes first occur at $1/2$, $1$ and $3/2$, matching the $\mathrm{SCP}$s of Eq.~\eqref{eq:xvcxvcxvcD}.}
 \label{eta}
\end{figure*}

\noindent \textbf{Informatics}\\
Let us now consider some measures typically used in quantum information science -- such as purity, linear entropy, concurrence and negativity -- and the impact upon them of the special points of interest listed in Eq.~\eqref{eq:xvcxvcxvc}, be they exceptional or otherwise. In particular, we are interested in the response of the system in the steady state. Notably, the steady-state entanglement properties of two coupled qubits is known to be important for quantum information processing and device design~\cite{Plenio2002, Hartmann2006, Valle2011,Gonzalez2011,Decordi2020}.

The measure $\mathrm{EPR}$ frequently arises in quantum metrology, where the inequality $\mathrm{EPR} > 0$ suggests entanglement~\cite{Pudlik2013}. Considering the steady state with the asymptotic time $t \to \infty$, and coming directly from the definition of $\mathrm{EPR} = \lim_{t\to\infty} ( \langle \sigma_1^\dagger\sigma_1\rangle \langle \sigma_2^\dagger\sigma_2\rangle - \langle \sigma_1^\dagger\sigma_1 \sigma_2^\dagger\sigma_2\rangle )$, we obtain this key measure as~\cite{Supplementary}
\begin{equation}
\label{eq:frt}
\mathrm{EPR} =  \left( \frac{s}{s+1} \frac{2g\kappa}{ 4 g^2 + s\kappa^2} \right)^2,
\end{equation}
which exhibits the lower bound of $0$ for both weak and strong coupling ($g \ll \kappa$ and $g \gg \kappa$ respectively), and the upper bound of $s/[2(1+s)]^2$ at the certain coupling strength $g = \sqrt{s}\kappa/2$. We plot the $\mathrm{EPR}$ in Fig.~\ref{eta}~(a) as a function of $g$, showcasing the stronger $\mathrm{EPR}$ entanglement for moderate coupling values and how it vanishes for extremely weak or strong couplings, such that no exceptional or critical points influence this simple measure.

The linear entropy $\mathcal{S}$ is a common measure of the mixedness of a quantum state $\rho$. This quantity may be defined by $\mathcal{S} = (4/3) [1 - \Tr (\rho^2) ] $, where the prefactor of $4/3$ ensures that the bounds fulfil $S \in [0, 1]$. The lower bound corresponds to a pure state and the upper bound to a maximally mixed state. In the steady state, the linear entropy $\mathcal{S}$ for the unbalanced qubit system reads~\cite{Supplementary}
\begin{align}
\label{eq:lin}
\mathcal{S} =& \frac{32}{3} \frac{s}{(s+1)^4} \left( \frac{g}{ 4 g^2 + s \kappa^2} \right)^2  \nonumber \\
&\times \bigg\{ (1+s^2)(1+s)^2 \kappa^2 + 8 (s^2+s+1) g^2 \bigg\}.
\end{align}
We plot the linear entropy in Fig.~\ref{eta}~(b) as a function of the coupling $g$, which highlights the total purity of the state with small couplings $g \ll \kappa$ as may be expected. The monotonic rise in mixedness of the state with increasing coupling $g$ eventually plateaus at a value given by $\mathcal{S} \simeq 16 s (1+s+s^2)/[3 (1+s)^4]$. Hence, only with balanced coupling ($s=1$) does a maximally mixed state arise with $\mathcal{S} = 1$ identically.

The concurrence $\mathcal{C}$, as developed by Wooters~\cite{Hill1997, Wootters1998}, is a celebrated entanglement measure of bipartite mixed states. Zero concurrence is associated with a separable state, while nonzero concurrence measures the degree of entanglement of the state (up to a maximum of unity for maximum entanglement). We consider the steady state concurrence $\mathcal{C}$ as a function of the coupling strength $g$ in Fig.~\ref{eta}~(c) for the cases of $s = \{ 1/2, 1, 2 \}$. Strikingly, there are critical points at which the steady state concurrence suddenly drops to zero such that the state becomes separable --  we calls these points separable critical points or $\mathrm{SCP}$s [cf. Eq.~\eqref{eq:xvcxvcxvcD}]. Otherwise, in the region of nonzero entanglement one finds (for the balanced case of $s=1$) the maximal obtainable concurrence
\begin{equation}
\label{eq:conc}
\text{max} \{ \mathcal{C} \}  = \frac{1}{1+\sqrt{5}} \simeq 0.309...,
\end{equation}
which occurs with the moderate coupling $g = \kappa/(1+\sqrt{5}) \simeq 0.309 \kappa$, and which is slightly higher than the maximum values of $\mathcal{C}$ for the unbalanced arrangements of the qubits. The entanglement-disentanglement transitions in Fig.~\ref{eta}~(c) are wholly governed by Eq.~\eqref{eq:xvcxvcxvcD}, which is a useful quantum point of interest in addition to the exceptional points of the system.
\\

\begin{figure*}[tb]
 \includegraphics[width=1.0\linewidth]{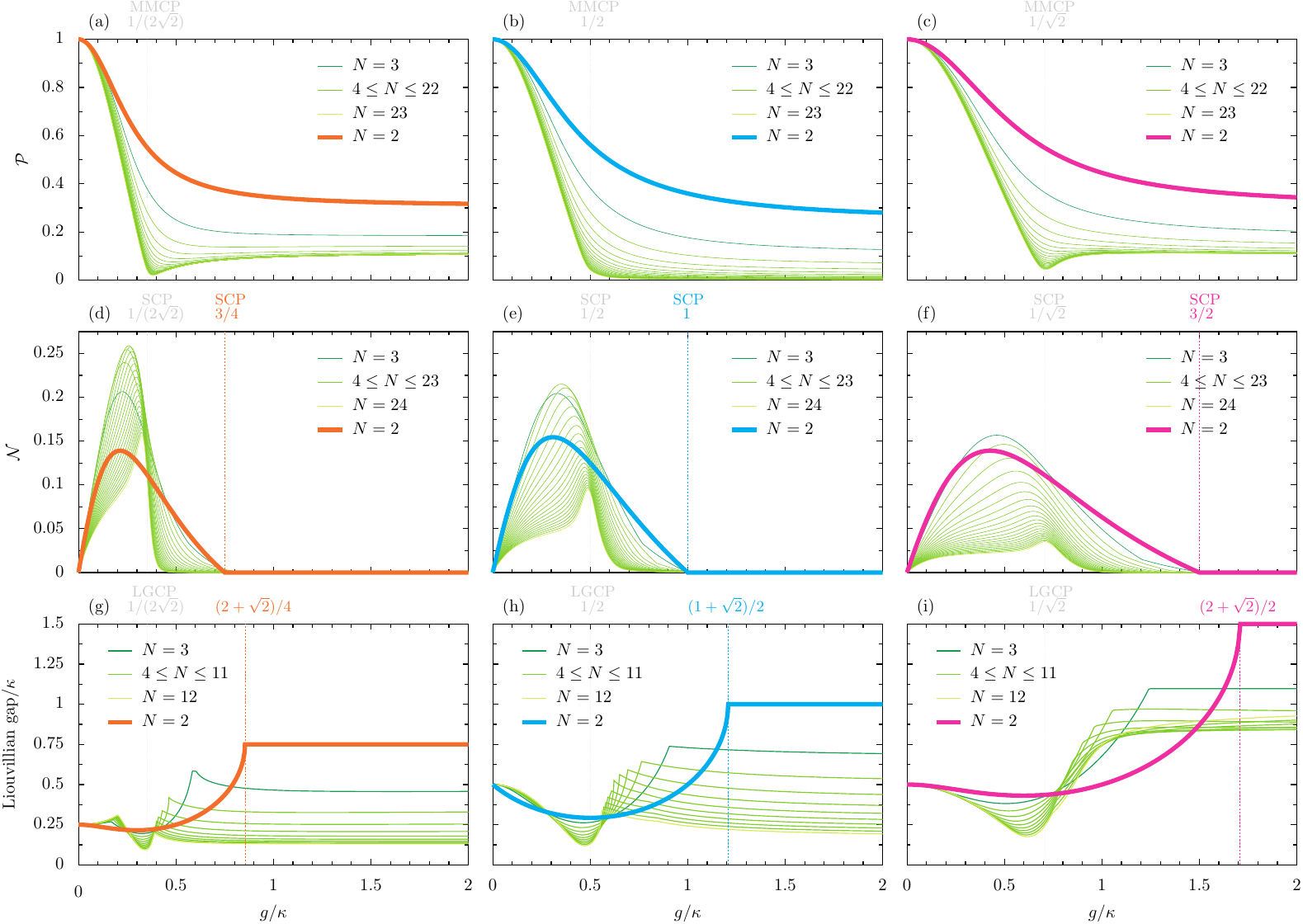}
 \caption{ \textbf{The unbalanced dimer beyond the two-level limit.} We consider the steady state ($t\to\infty$) limit, and the two-level system results are denoted by thick orange, cyan and pink lines for the cases of $s = \{ 1/2, 1, 2 \}$ respectively. Also shown are the results for a pair of truncated quantum harmonic oscillators: the number of levels of each oscillator is $N$, so that $N = 2$ recovers the two-level system results while untruncated oscillators are approached in the $N \gg 2$ limit (thin green lines). Panels (a, b, c): the purity $\mathcal{P}$ as a function of the coupling strength $g$ (in units of $\kappa$). Dotted grey lines: the values of interest as $N \gg 2$ are $1/(2\sqrt{2}) \simeq 0.354$, $1/2$ and $1/\sqrt{2} \simeq 0.707$ for panels (a), (b) and (c) respectively, aligning with the $\mathrm{MMCP}$ for the coupled oscillators [cf. Eq.~\eqref{eq:xvcxvbnmcxvcC}]. Panels (d, e, f):  as for the upper panels, but instead showing the negativity $\mathcal{N}$ as a function of $g$. Dotted grey lines: the values of interest as $N \gg 2$ are $1/(2\sqrt{2}) \simeq 0.354$, $1/2$ and $1/\sqrt{2} \simeq 0.707$ for panels (d), (e) and (f) respectively, aligning with the $\mathrm{SCP}$ for the coupled oscillators [cf. Eq.~\eqref{eq:xvcxvbnmcxvcC}]. Dashed grey lines: the negativity zeroes along the top row of panels occur at $3/4$, $1$ and $3/2$ respectively for $N=2$, which are the qubit $\mathrm{SCP}$s [cf. Eq.~\eqref{eq:xvcxvcxvcD}]. Panels (g, h, i): as for the upper panels, but instead showing the Liouvillian gap as a function of $g$. Dotted grey lines: the values of interest as $N \gg 2$ are $1/(2\sqrt{2}) \simeq 0.354$, $1/2$ and $1/\sqrt{2} \simeq 0.707$ for panels (g), (h) and (i) respectively, these points are $\mathrm{LGCP}$s for the coupled oscillators [cf. Eq.~\eqref{eq:xvcxvbnmcxvcC}]. Dashed grey lines: the Liouvillian gap plateaus along the bottom row of panels occur at $(\sqrt{2}+2)/4 \simeq 0.85$, $(\sqrt{2}+1)/2 \simeq 1.21$ and $(\sqrt{2}+2)/\sqrt{2} \simeq 1.71$ respectively for $N=2$, aligning with the largest $\mathrm{type-I~MEP}$ for the coupled qubits [cf. Eq.~\eqref{eq:xvcxvcxvcB}].}
 \label{negat}
\end{figure*}

%===========================================================================
%===========================================================================
%===========================================================================
%===========================================================================
%===========================================================================
%===========================================================================
%===========================================================================
%===========================================================================

\noindent \textbf{Thermodynamic limit}\\
The results for the considered coupled qubits [cf. Eq.~\eqref{eq:Hammy}] may be approached from the analogous model of two coupled quantum harmonic oscillators by truncating the number of energy levels $N$ of each oscillator. In the untruncated limit ($N \to \infty$), the pure coupled oscillator system admits the following quantum points of interest~\cite{Supplementary}
\begin{subequations}
\label{eq:xvcxvdfdfdfcxvc}
\begin{align}
    g &=\left( \frac{s+1}{4} \right) \kappa, && 
    \begin{cases}
  \mathrm{type-I~MEP}, \\
 \mathrm{type-II~MEP},
\end{cases} \label{eq:xbmnvcxvcxvcB} \\
  g &= \frac{\sqrt{s}}{2} \kappa, &&
    \begin{cases}
  \mathrm{MMCP}, \\
    \mathrm{SCP}, \\
 \mathrm{LGCP}.
\end{cases} \label{eq:xvcxvbnmcxvcC} 
  \end{align}
\end{subequations}
Notably, the $\mathrm{type-I}$ and $\mathrm{type-II}$ $\mathrm{MEP}$s coincide for this ($N \to \infty$ level) linear model [Eq.~\eqref{eq:xbmnvcxvcxvcB}], in stark contrast to the qubit case with $N = 2$ levels [Eq.~\eqref{eq:xvcxvcxvcA} and Eq.~\eqref{eq:xvcxvcxvcB}]. Interestingly, there is also a coexistence of the separable critical point ($\mathrm{SCP}$) and the Liouvillian gap critical point ($ \mathrm{LGCP}$), defined as the point in parameter space at which the Liouvillian gap closes, which can indeed occur in the thermodynamic limit of $N \to \infty$. Additionally, there is a further coincidence in Eq.~\eqref{eq:xvcxvbnmcxvcC} with the maximally mixed critical point ($ \mathrm{MMCP}$), which is associated with purity of the state becoming zero.

In what follows, we consider for a handful of illuminating quantum steady state properties in the extreme limits of $N = 2$ for coupled qubits, $N \to \infty$ for coupled quantum harmonic oscillators, and otherwise for some values of $N$ in between these instructive limits.

Let us first consider the purity $\mathcal{P}$ of the system, a measure of the mixedness of the state $\rho$ as defined by $\mathcal{P} = \Tr (\rho^2)$, so that $\mathcal{P} = 1$ corresponds to a pure state and $\mathcal{P} = 1/D$ to a maximally mixed state (here $D$ is the dimension of the Hilbert space in question). The top row of Fig.~\ref{negat} displays the steady state purity $\mathcal{P}$ as a function of the coupling strength $g$, for the cases of $s = \{ 1/2, 1, 2 \}$ in panels (a), (b) and (c) respectively. The qubit ($N=2$) results are given by the thick colored lines in the figure, and are described by the exact expression
\begin{equation}
\label{eq:linsdfsdfs}
\mathcal{P} = \frac{ 16g^2 \left[ s^2 (1+s)^2 \kappa^2 +  (1+s^2)^2 g^2 \right] + s^2 (1+s)^4 \kappa^4 }{(1+s)^4 \left( 4g^2 + s \kappa^2  \right)^2},
\end{equation}
which is necessarily unity with vanishing coupling $g$. The truncated oscillator results are represented by the thin green lines in Fig.~\ref{negat}~(a, b, c), and for greater visibility the edge cases of $N = 3$ and $N = 20$ are given by dark green and light green lines respectively. Notably, with increasing the number of levels $N$ the purity $\mathcal{P}$ tends towards zero -- suggesting maximal mixedness -- at the $\mathrm{MMCP}$ defined by Eq.~\eqref{eq:xvcxvbnmcxvcC}, which is marked by the dotted grey lines in the figure.

The negativity $\mathcal{N}$ is a popular quantifier of how much entanglement is contained within a quantum state $\rho$~\cite{Zyczkowski1998,Vidal2002}. This quantity is zero for separable states and nonzero for entangled states. Defined as the absolute sum of the negative eigenvalues of the partial transpose of the state $\rho$, it may be computed exactly in the steady state and for the $N = 2$ level case as~\cite{Supplementary}
\begin{align}
\label{eq:neg}
\mathcal{N} =& \frac{ 2 g }{4g^2+s\kappa^2} \bigg\{ \sqrt{\left( s-1\right)^2 g^2 + s^2 \kappa^2} - \left( \frac{s^2+1}{s+1} \right) g \bigg\} \nonumber \\
 &\times \frac{1}{s+1}  \Theta \left( \tfrac{s+1}{2}  \kappa - g \right), 
\end{align}
where $\Theta (x)$ is the Heaviside theta function. The presence of the step function within Eq.~\eqref{eq:neg} highlights the hard border between entangled and unentangled states at the critical points defined by Eq.~\eqref{eq:xvcxvcxvcD} for the qubit ($N = 2$) case. For example, within the balanced arrangement of the qubit dimer ($s=1$) we find the maximal negativity $\text{max} \{ \mathcal{N} \}  = 1 / (2+2\sqrt{5})\simeq 0.155$ occurs at $g = \kappa / (1+\sqrt{5})\simeq 0.309$, while the $\mathrm{SCP}$ may be found at $g = \kappa/2$. This circumstance is plotted as the thick cyan line in Fig.~\ref{negat}~(e), while the panels (d) and (f) show the equivalent results for the unbalanced cases of $s=1/2$ (thick orange line) and $s=2$ (thick pink line) respectively, where the qubit $\mathrm{SCP}$s are denoted by the dashed colored lines. The cases of truncated oscillators are shown in the second row of Fig.~\ref{negat} by thin green lines. Notably, in the thermodynamic limit $N \to \infty$ the coupled oscillator negativities $\mathcal{N}$ in Fig.~\ref{negat}~(d, e, f) seemingly go to zero at the $\mathrm{SCP}$s predicted by Eq.~\eqref{eq:xvcxvbnmcxvcC}, which are marked by the dotted grey lines in the figure. This statement is most easily seen in panel (d), and more supporting evidence for this claim is presented in \cite{Supplementary}.

The Liouvillian spectral gap is defined from the complex eigenvalue of the Liouvillian with the smallest real part (after removing any zero eigenvalues from consideration)~\cite{Kessler2012, Cai2013, Minganti2018}. The closing of the Liouvillian gap, which is possible in the thermodynamic limit $N \to \infty$, may occur at a critical value of some system parameter. This gap closing phenomena is associated with a dissipative phase transition, which have already been measured in a handful of modern photonic architectures~\cite{Fitzpatrick2017, Rodriguez2017, Fink2018}. We plot the Liouvillian gap as a function of the coupling strength $g$ in Fig.~\ref{negat}~(g, h, i) for three different values of $s$. The coupled qubit ($N=2$) results are denoted by thick colored lines and cannot lead to a closing of the Liouvillian gap due to the finite size of the Hilbert space, however the plateau feature arises due to the largest $\mathrm{type-I~MEP}$ [cf. Eq.~\eqref{eq:xvcxvcxvcB}]. Interestingly, the truncated oscillator results (thin green lines) suggest a closing of the gap in the large $N$ limit exactly at the $\mathrm{LGCP}$ of Eq.~\eqref{eq:xvcxvbnmcxvcC}. This trend of gap closing is best seen in panel (g), and more supporting evidence for this claim is given in \cite{Supplementary}.

Overall, by truncating the linear harmonic oscillator model to recover the qubits results, one notices the evolution of the $\mathrm{SCP}$s in the middle row of Fig.~\ref{negat}, describing the movement in $g/\kappa$-space of the entanglement-disentanglement transitions in the steady state. The consideration of a thermodynamic limit with the coupled oscillators also revealed the critical points that we have dubbed $\mathrm{MMCP}$ and $\mathrm{LGCP}$, governing the mixedness zero and the dissipative phase transition in the system using Eq.~\eqref{eq:xvcxvbnmcxvcC}, which complement the more celebrated exceptional points of the oscillator model [cf. Eq.~\eqref{eq:xbmnvcxvcxvcB}].
\\

%===========================================================================
%===========================================================================
%===========================================================================
%===========================================================================
%===========================================================================
%===========================================================================
%===========================================================================
%===========================================================================
\noindent \textbf{Discussion}\\
In conclusion, we have considered a simple quantum optical model of two coupled qubits experiencing both loss and gain. In analogy to the (essentially semiclassical) exceptional points found when using non-Hermitian Hamiltonians to model optical systems, here we have elucidated the many flavours of quantum points of interest -- both exceptional and critical -- which should appear in quantum optical systems describable within an open quantum systems approach. In particular, we have suggested that a change in the number of peaks in the optical spectrum can be a neat signifier of movement through a certain flavour of first moment exceptional point, while steady state entanglement measures abruptly becoming zero are associated with passing past a certain critical point. Following a series of recent pioneering experiments in quantum non-Hermitian physics -- primarily exploiting superconducting qubits~\cite{Naghiloo2019, Partanen2019, Dogra2021, Chen2021, Chen2022,Abbasi2022, Zhang2022, Bu2023, Han2022, Han2022b, Liang2023, Li2023, Quinn2023} -- we hope that our theoretical results can stimulate further experimental work in the study of quantum points of interest, broadly interpreted, and their distinctive consequences. 
\\

%===========================================================================
%===========================================================================
%===========================================================================
%===========================================================================
%===========================================================================
%===========================================================================
%===========================================================================
%===========================================================================
\noindent \textbf{Acknowledgments}\\
\textit{Funding}: CAD is supported by the Royal Society via a University Research Fellowship (URF\slash R1\slash 201158). OIRF is funded by the EPSRC via the Maths DTP 2021-22 University of Exeter (EP/W523859/1). \textit{Discussions}: We thank V.~A.~Saroka for fruitful discussions. \textit{Data and materials availability}: All data is available within the manuscript and the Supplementary Material.
\\

%===========================================================================
%===========================================================================
%===========================================================================
%===========================================================================
%===========================================================================
%===========================================================================
%===========================================================================
%===========================================================================

\noindent
\textbf{ORCID}\\
C. A. Downing: \href{https://orcid.org/0000-0002-0058-9746}{0000-0002-0058-9746}.
\\

%===========================================================================
%===========================================================================
%===========================================================================
%===========================================================================
%===========================================================================
%===========================================================================
%===========================================================================
%===========================================================================

\noindent
\textbf{Supplementary Material}\\
This paper has some Supplementary Material available alongside it. The Supplementary Material includes details of the theoretical framework underpinning our study of a pair of driven-dissipative two-level systems, as well as a brief review of the complementary situation with quantum harmonic oscillators replacing the two-level systems.
\\

%===========================================================================
%===========================================================================
%===========================================================================
%===========================================================================
%===========================================================================
%===========================================================================
%===========================================================================
%===========================================================================

%===========================================================================
%===========================================================================
%===========================================================================
%===========================================================================
%===========================================================================
%===========================================================================
%===========================================================================
%===========================================================================


\begin{thebibliography}{100}

%===========================================================================
%===========================================================================
%===========================================================================
%===========================================================================
%===========================================================================
%===========================================================================
%===========================================================================
%===========================================================================

\bibitem{Bender2007}
C.~M.~Bender,
Making sense of non-Hermitian Hamiltonians,
\href{https://doi.org/10.1088/0034-4885/70/6/R03}
{Rep. Prog. Phys. \textbf{70}, 947 (2007)}.

\bibitem{Moiseyev2011}
N.~Moiseyev,
\textit{Non-Hermitian Quantum Mechanics} (Cambridge University Press, Cambridge, 2011).

\bibitem{Ashida2020}
Y.~Ashida, Z.~Gong and M.~Ueda,
Non-Hermitian physics,
\href{https://doi.org/10.1080/00018732.2021.1876991}
{Adv. Phys. \textbf{69}, 249 (2020)}.


\bibitem{Konotop2016}
V.~V.~Konotop, J.~Yang and D.~A.~Zezyulin,
Nonlinear waves in $\mathcal{PT}$-symmetric systems,
\href{https://doi.org/10.1103/RevModPhys.88.035002}
{Rev. Mod. Phys. \textbf{88}, 035002 (2016)}.

\bibitem{Ganainy2018}
R.~El-Ganainy, K.~G.~Makris, M.~Khajavikhan, Z.~H.~Musslimani, S.~Rotter and D.~N.~Christodoulides,
Non-Hermitian physics and $\mathcal{PT}$ symmetry,
\href{https://doi.org/10.1038/nphys4323}
{Nature Phys. \textbf{14}, 11 (2018)}.


\bibitem{Miri2019}
M.-A.~Miri and A.~Alu,
Exceptional points in optics and photonics,
\href{https://doi.org/10.1126/science.aar7709}
{Science \textbf{363}, 6422 (2019)}.

\bibitem{Ozdemir2019}
S.~K.~Ozdemir, S.~Rotter, F.~Nori and L.~Yang,
Parity-time symmetry and exceptional points in photonics,
\href{https://doi.org/10.1038/s41563-019-0304-9}
{Nat. Mater. \textbf{18}, 783 (2019)}.

\bibitem{Berry2003}
M.~V.~Berry,
Physics of nonhermitian degeneracies,
\href{https://doi.org/10.1023/B:CJOP.0000044002.05657.04}
{Czech. J. Phys. \textbf{54}, 1039 (2004)}.

\bibitem{Heiss2003}
W.~D.~Heiss,
Exceptional points -- their universal occurrence and their physical significance,
\href{https://doi.org/10.1023/B:CJOP.0000044009.17264.dc}
{Czech. J. Phys. \textbf{54}, 1091 (2004)}.

\bibitem{Ruter2010}
C.~E.~Ruter, K.~G.~Makris, R.~El-Ganainy, D.~N.~Christodoulides, M.~Segev and D.~Kip, 
Observation of parity-time symmetry in optics,
\href{https://doi.org/10.1038/nphys1515}
{Nature Phys. \textbf{6}, 192 (2010)}.

\bibitem{Feng2014}
L.~Feng, Z.~J.~Wong, R.-M.~Ma, Y.~ Wang and X.~Zhang, 
Single-mode laser by parity-time symmetry breaking,
\href{https://doi.org/10.1126/science.1258479}
{Science \textbf{346}, 972 (2014)}.

\bibitem{Hodaei2014}
H.~Hodaei, M.-A.~Miri, M.~Heinrich, D.~N.~Christodoulides and M.~Khajavikhan, 
Parity-time-symmetric microring lasers,
\href{https://doi.org/10.1126/science.1258480}
{Science \textbf{346}, 975 (2014)}.

\bibitem{Hodaei2017}
H.~Hodaei, A.~U.~Hassan, S.~Wittek, H.~Garcia-Gracia, R.~El-Ganainy, D.~N.~Christodoulides and M.~Khajavikhan, 
Enhanced sensitivity at higher-order exceptional points,
\href{https://doi.org/10.1038/nature23280}
{Nature \textbf{548}, 187 (2017)}.

\bibitem{Wu2019}
Y.~Wu, W.~Liu, J.~Geng, X.~Song, X.~Ye, C.-K.~Duan, X.~Rong and J.~Du, 
Observation of parity-time symmetry breaking in a single-spin system,
\href{https://doi.org/10.1126/science.aaw8205}
{Science \textbf{364}, 878 (2019)}.

\bibitem{Xia2021}
S.~Xia, D.~Kaltsas, D.~Song, I.~Komis, J.~Xu, A.~Szameit, H.~Buljan, K.~G. Makris and Z.~Chen,
Nonlinear tuning of $\mathcal{PT}$ symmetry and non-Hermitian topological states,
\href{https://doi.org/10.1126/science.abf6873}
{Science \textbf{372}, 72 (2021)}.

\bibitem{Sweeney2021}
C.~Wang, W.~R.~Sweeney, A.~D.~Stone and L.~Yang,
Coherent perfect absorption at an exceptional point,
\href{https://doi.org/10.1126/science.abj1028}
{Science \textbf{373}, 1261 (2021)}.

\bibitem{Ergoktas2022}
M.~S.~Ergoktas, S.~Soleymani, N.~Kakenov, K.~Wang, T.~Smith, G.~Bakan, S.~Balci, A.~Principi, K.~Novoselov, S.~K.~Ozdemir and C.~Kocabas,
Topological engineering of terahertz light using electrically tunable exceptional point singularities,
\href{https://doi.org/10.1126/science.abn6528}
{Science \textbf{376}, 184 (2022)}.

\bibitem{Prosen2012}
T.~Prosen,
$\mathcal{PT}$-Symmetric Quantum Liouvillean Dynamics,
\href{https://doi.org/10.1103/PhysRevLett.109.090404}
{Phys. Rev. Lett. \textbf{109}, 090404 (2012)}.

\bibitem{Prosen2012b}
T.~Prosen,
Generic examples of $\mathcal{PT}$-symmetric qubit (spin-1/2) Liouvillian dynamics,
\href{https://doi.org/10.1103/PhysRevA.86.044103}
{Phys. Rev. A \textbf{86}, 044103 (2012)}.


\bibitem{Shallem2015}
M.~Am-Shallem, R.~Kosloff and N.~Moiseyev,
Exceptional points for parameter estimation in open quantum systems: analysis of the Bloch equations,
\href{https://doi.org/10.1088/1367-2630/17/11/113036}
{New J. Phys. 17 113036 \textbf{17}, 113036 (2012)}.


\bibitem{Kepesidis2016}
K.~V.~Kepesidis, T.~J.~Milburn, J.~Huber, K.~G.~Makris, S.~Rotter and P.~Rabl,
$\mathcal{PT}$-symmetry breaking in the steady state of microscopic gain–loss systems,
\href{https://doi.org/10.1088/1367-2630/18/9/095003}
{New J. Phys. \textbf{18}, 095003 (2016)}.


\bibitem{Minganti2019}
F.~Minganti, A.~Miranowicz, R.~W.~Chhajlany and F.~Nori,
Quantum exceptional points of non-Hermitian Hamiltonians and Liouvillians: The effects of quantum jumps,
\href{https://doi.org/10.1103/PhysRevA.100.062131}
{Phys. Rev. A \textbf{100}, 062131 (2019)}.


\bibitem{Jaramillo2020}
B.~Jaramillo~Avila, C.~Ventura-Velazquez, R.~de~J.~Leon-Montiel, Y.~N.~Joglekar and B.~M.~Rodriguez-Lara,
$\mathcal{PT}$-symmetry from Lindblad dynamics in a linearized optomechanical system,
\href{https://doi.org/10.1038/s41598-020-58582-7}
{Sci. Rep. \textbf{10}, 1761 (2020)}.


\bibitem{Arkhipov2020}
I.~I.~Arkhipov, A.~Miranowicz, F.~Minganti and F.~Nori,
Quantum and semiclassical exceptional points of a linear system of coupled cavities with losses and gain within the Scully-Lamb laser theory,
\href{https://doi.org/10.1103/PhysRevA.101.013812}
{Phys. Rev. A \textbf{101}, 013812 (2020)}.

\bibitem{Downing2021}
C.~A.~Downing and V.~A.~Saroka,
Exceptional points in oligomer chains,
\href{https://doi.org/10.1038/s42005-021-00757-3}
{Commun. Phys. \textbf{4}, 254 (2021)}.

\bibitem{Huber2020}
J.~Huber, P.~Kirton, S.~Rotter and P.~Rabl,
Emergence of $\mathcal{PT}$-symmetry breaking in open quantum systems,
\href{http://dx.doi.org/10.21468/SciPostPhys.9.4.052}
{SciPost Phys. \textbf{9}, 052 (2020)}.


\bibitem{Khandelwal2021}
S.~Khandelwal, N.~Brunner and G.~Haack,
Signatures of Liouvillian exceptional points in a quantum thermal machine,
\href{https://doi.org/10.1103/PRXQuantum.2.040346}
{PRX Quantum \textbf{2}, 040346 (2021)}.

\bibitem{Lopez2020}
C.~A.~Downing, J.~C.~López~Carreño, A.~I.~Fernández-Domínguez and E.~del~Valle,
Asymmetric coupling between two quantum emitters,
\href{https://doi.org/10.1103/PhysRevA.102.013723}
{Phys. Rev. A \textbf{102}, 013723 (2020)}.

\bibitem{Supplementary}
Please see the Supplementary Material for the background theory supporting the results reported in the main text. It includes a discussion of the refilling terms in the quantum master equation of Eq.~\eqref{eqapp:massdsdsdter}, and the physics of the non-Hermitian Hamiltonian limit of the coupled qubit model.

\bibitem{Allen1975}
L.~Allen and J.~H.~Eberly,
\textit{Optical Resonance and Two-Level Atoms} (Wiley, New York, 1975).


\bibitem{Zueco2021}
C.~A.~Downing and D.~Zueco,
Non-reciprocal population dynamics in a quantum trimer,
\href{https://doi.org/10.1098/rspa.2021.0507}
{Proc. R. Soc. A \textbf{477}, 20210507 (2021)}.


\bibitem{Gardiner2014}
C.~Gardiner and P.~Zoller,
\textit{The Quantum World of Ultra-Cold Atoms and Light, Book I: Foundations of Quantum Optics} (Imperial College Press, London, 2014).

\bibitem{Reiter2014}
T.~E.~Lee, F.~Reiter and N.~Moiseyev,
Entanglement and spin squeezing in non-Hermitian phase transitions,
\href{https://doi.org/10.1103/PhysRevLett.113.250401}
{Phys. Rev. Lett. \textbf{113}, 250401 (2014)}.



\bibitem{Ramirez2019}
R.~Ramirez and M.~Reboiro,
Optimal spin squeezed steady state induced by the dynamics of non-Hermitian Hamiltonians,
\href{https://doi.org/10.1088/1402-4896/ab0fc0}
{Phys. Scr. \textbf{94}, 085220 (2019)}.


\bibitem{Gardiner1988}
C.~W.~Gardiner,
Quantum noise and quantum Langevin equations,
\href{https://doi.org/10.1147/rd.321.0127}
{IBM J. Res. Develop. \textbf{32}, 127 (1988)}.



\bibitem{Bender1998}
C.~M.~Bender and S.~Boettcher,
Real spectra in non-Hermitian Hamiltonians having $\mathcal{PT}$ symmetry,
\href{https://doi.org/10.1103/PhysRevLett.80.5243}
{Phys. Rev. Lett. \textbf{80}, 5243 (1998)}.


\bibitem{Bender2018}
C.~M.~Bender,
\textit{$\mathcal{PT}$ Symmetry: in Quantum and Classical Physics} (World Scientific, Singapore, 2018).


\bibitem{Sturges2022}
C.~A.~Downing and T.~J.~Sturges,
Directionality between driven-dissipative resonators,
\href{https://doi.org/10.1209/0295-5075/ac9ad6}
{EPL \textbf{140}, 35001 (2022)}.


\bibitem{Manzano2012}
D.~Manzano, M.~Tiersch, A.~Asadian and H.~J.~Briegel,
Quantum transport efficiency and Fourier's law,
\href{https://doi.org/10.1103/PhysRevE.86.061118}
{Phys. Rev. E \textbf{86}, 061118 (2012)}.


\bibitem{Valle2010}
E.~del~Valle,
Strong and weak coupling of two coupled qubits,
\href{https://doi.org/10.1103/PhysRevA.81.053811}
{Phys. Rev. A \textbf{81}, 053811 (2010)}.

\bibitem{Downing2023}
C. A. Downing, E. del Valle, and A. I. Fernández-Domínguez,
Resonance fluorescence of two asymmetrically pumped and coupled two-level systems,
\href{https://doi.org/10.1103/PhysRevA.107.023717}
{Phys. Rev. A \textbf{107}, 023717 (2023)}.



\bibitem{Pudlik2013}
T.~Pudlik, H.~Hennig, D.~Witthaut and D.~K.~Campbell,
Dynamics of entanglement in a dissipative Bose-Hubbard dimer,
\href{https://doi.org/10.1103/PhysRevA.88.063606}
{Phys. Rev. A \textbf{88}, 063606 (2013)}.





\bibitem{Plenio2002}
M.~B.~Plenio and S.~F.~Huelga,
Entangled light from white noise,
\href{https://doi.org/10.1103/PhysRevLett.88.197901}
{Phys. Rev. Lett. \textbf{88}, 197901 (2002)}.

\bibitem{Hartmann2006}
L.~Hartmann, W.~Dur and H.-J.~Briegel,
Steady-state entanglement in open and noisy quantum systems,
\href{https://doi.org/10.1103/PhysRevA.74.052304}
{Phys. Rev. A \textbf{74}, 052304 (2006)}.

\bibitem{Valle2011}
E.~del~Valle,
Steady-state entanglement of two coupled qubits,
\href{https://doi.org/10.1364/JOSAB.28.000228}
{J. Opt. Soc. Am. B \textbf{28}, 228 (2011)}.

\bibitem{Gonzalez2011}
A.~Gonzalez-Tudela, D.~Martin-Cano, E.~Moreno, L.~Martin-Moreno, C.~Tejedor and F.~J.~Garcia-Vidal,
Entanglement of two qubits mediated by one-dimensional plasmonic waveguides,
\href{https://doi.org/10.1103/PhysRevLett.106.020501}
{Phys. Rev. Lett. \textbf{106}, 020501 (2011)}.


\bibitem{Decordi2020}
G.~L.~Decordi and A.~Vidiella-Barranco,
Sudden death of entanglement induced by a minimal thermal environment,
\href{https://doi.org/10.1016/j.optcom.2020.126233}
{Opt. Commun. \textbf{475}, 126233 (2020)}.


\bibitem{Hill1997}
S.~A.~Hill and W.~K.~Wootters,
Entanglement of a pair of quantum bits,
\href{https://doi.org/10.1103/PhysRevLett.78.5022}
{Phys. Rev. Lett. \textbf{78}, 5022 (1997)}.

\bibitem{Wootters1998}
W.~K.~Wootters,
Entanglement of formation of an arbitrary state of two qubits,
\href{https://doi.org/10.1103/PhysRevLett.80.2245}
{Phys. Rev. Lett. \textbf{80}, 2245 (1998)}.



\bibitem{Zyczkowski1998}
K.~Zyczkowski, P.~Horodecki, A.~Sanpera and M.~Lewenstein,
Volume of the set of separable states,
\href{https://doi.org/10.1103/PhysRevA.58.883}
{Phys. Rev. A \textbf{58}, 883 (1998)}.

\bibitem{Vidal2002}
G.~Vidal and R.~F.~Werner,
Computable measure of entanglement,
\href{https://doi.org/10.1103/PhysRevA.65.032314}
{Phys. Rev. A \textbf{65}, 032314 (2002)}.

\bibitem{Kessler2012}
E.~M.~Kessler, G.~Giedke, A.~Imamoglu, S.~F.~Yelin, M.~D.~Lukin and J.~I.~Cirac, 
Dissipative phase transition in a central spin system,
\href{https://doi.org/10.1103/PhysRevA.86.012116}
{Phys. Rev. A \textbf{86}, 012116 (2012)}.

\bibitem{Cai2013}
Z.~Cai and T.~Barthel, 
Algebraic versus exponential decoherence in dissipative many-particle systems,
\href{https://doi.org/10.1103/PhysRevLett.111.150403}
{Phys. Rev. Lett. \textbf{111}, 150403 (2013)}.

\bibitem{Minganti2018}
F.~Minganti, A.~Biella, N.~Bartolo and C.~Ciuti, 
Spectral theory of Liouvillians for dissipative phase transitions,
\href{https://doi.org/10.1103/PhysRevA.98.042118}
{Phys. Rev. A \textbf{98}, 042118 (2018)}.


\bibitem{Fitzpatrick2017}
M.~Fitzpatrick, N.~M.~Sundaresan, A.~C.~Y.~Li, J.~Koch and A.~A.~Houck, 
Observation of a dissipative phase transition in a one-dimensional circuit QED lattice,
\href{https://doi.org/10.1103/PhysRevX.7.011016}
{Phys. Rev. X \textbf{7}, 011016 (2017)}.


\bibitem{Rodriguez2017}
S.~R.~K.~Rodriguez, W.~Casteels, F.~Storme, N.~C.~Zambon, I.~Sagnes, L.~Le~Gratiet, E.~Galopin, A.~Lemaitre, A.~Amo, C.~Ciuti and J.~Bloch, 
Probing a dissipative phase transition via dynamical optical hysteresis,
\href{https://doi.org/10.1103/PhysRevLett.118.247402}
{Phys. Rev. Lett. \textbf{118}, 247402 (2017)}.

\bibitem{Fink2018}
T.~Fink, A.~Schade, S.~Hofling, C.~Schneider and A.~Imamoglu, 
Signatures of a dissipative phase transition in photon correlation measurements,
\href{https://doi.org/10.1038/s41567-017-0020-9}
{Nat. Phys. \textbf{14}, 365 (2018)}.


\bibitem{Naghiloo2019}
M.~Naghiloo, M.~Abbasi, Y.~N.~Joglekar and K.~W.~Murch, 
Quantum state tomography across the exceptional point in a single dissipative qubit,
\href{https://doi.org/10.1038/s41567-019-0652-z}
{Nat. Phys. \textbf{15}, 1232 (2019)}.


\bibitem{Partanen2019}
M.~Partanen, J.~Goetz, K.~Y.~Tan, K.~Kohvakka, V.~Sevriuk, R.~E.~Lake, R.~Kokkoniemi, J.~Ikonen, D.~Hazra, A.~Makinen, E.~Hyyppa, L.~Gronberg, V.~Vesterinen, M.~Silver and M.~Mottonen,
Exceptional points in tunable superconducting resonators,
\href{https://doi.org/10.1103/PhysRevB.100.134505}
{Phys. Rev. B \textbf{100}, 134505 (2019)}.

\bibitem{Chen2021}
W.~Chen, M.~Abbasi, Y.~N.~Joglekar and K.~W.~Murch, 
Quantum jumps in the non-Hermitian dynamics of a superconducting qubit,
\href{https://doi.org/10.1103/PhysRevLett.127.140504}
{Phys. Rev. Lett. \textbf{127}, 140504 (2021)}.

\bibitem{Dogra2021}
S.~Dogra, A.~A.~Melnikov and G.~S.~Paraoanu,
Quantum simulation of parity–time symmetry breaking with a superconducting quantum processor,
\href{https://doi.org/10.1038/s42005-021-00534-2}
{Commun. Phys. \textbf{4}, 26 (2021)}.

\bibitem{Chen2022}
W.~Chen, M.~Abbasi, B.~Ha, S.~Erdamar, Y.~N.~Joglekar and K.~W.~Murch,
Decoherence induced exceptional points in a dissipative superconducting qubit,
\href{https://doi.org/10.1103/PhysRevLett.128.110402}
{Phys. Rev. Lett. \textbf{128}, 110402 (2022)}.

\bibitem{Abbasi2022}
M.~Abbasi, W.~Chen, M.~Naghiloo, Y.~N.~Joglekar and K.~W.~Murch,
Topological quantum state control through exceptional-point proximity,
\href{https://doi.org/10.1103/PhysRevLett.128.160401}
{Phys. Rev. Lett. \textbf{128}, 160401 (2022)}.

\bibitem{Zhang2022}
J.-W.~Zhang, J.-Q.~Zhang, G.-Y.~Ding, J.-C.~Li, J.-T.~Bu, B.~Wang, L.-L.~Yan, S.-L.~Su, L.~Chen, F.~Nori, S.~K.~Ozdemir, F.~Zhou, H.~Jing and M.~Feng,
Dynamical control of quantum heat engines using exceptional points,
\href{https://doi.org/10.1038/s41467-022-33667-1}
{Nat. Commun. \textbf{13}, 6225 (2022)}.

\bibitem{Bu2023}
J.~T.~Bu, J.~Q.~Zhang, G.~Y.~Ding, J.~C.~Li, J.~W.~Zhang, B.~Wang, W.~Q.~Ding, W.~F.~Yuan, L.~Chen, S.~K.~Ozdemir, F.~Zhou, H.~Jing and M.~Feng,
Enhancement of quantum heat engine by encircling a Liouvillian exceptional point,
\href{https://doi.org/10.1103/PhysRevLett.130.110402}
{Phys. Rev. Lett. \textbf{130}, 110402 (2023)}.

\bibitem{Han2022}
P.-R.~Han, F.~Wu, X.-J.~Huang, H.~Wu, Z.-B.~Yang, C.-L.~Zou, W.~Yi, M.~Zhang, H.~Li, K.~Xu, D.~Zheng, H.~Fan, J.~Wen and S.-B.~Zheng,
$\mathcal{PT}$ symmetry and $\mathcal{PT}$-enhanced quantum sensing in a spin-boson system,
\href{https://arxiv.org/abs/2210.04494v1}
{arXiv:2210.04494v1}.

\bibitem{Han2022b}
P.~Han, F.~Wu, X.~Huang, H.~Wu, C.-L.~Zou, W.~Yi, M.~Zhang, H.~Li, K.~Xu, D.~Zheng, H.~Fan, J.~Wen, Z.~Yang and S.~Zheng,
Exceptional entanglement transition and natural-dissipation-enhanced quantum sensing,
\href{https://arxiv.org/abs/2210.04494v2}
{arXiv:2210.04494v2}.

\bibitem{Liang2023}
C.~Liang, Y.~Tang, A.-N.~Xu and Y.-C.~Liu,
Observation of exceptional points in thermal atomic ensembles,
\href{https://arxiv.org/abs/2304.06985}
{arXiv:2304.06985}.

\bibitem{Li2023}
Z.-Z.~Li, W.~Chen, M.~Abbasi, K.~W.~Murch, K.~B.~Whaley, S.~K.~Ozdemir, F.~Zhou, H.~Jing and M.~Feng,
Speeding up entanglement generation by proximity to higher-order exceptional points,
\href{https://arxiv.org/abs/2210.05048}
{arXiv:2210.05048}.

\bibitem{Quinn2023}
A.~Quinn, J.~Metzner, J.~E.~Muldoon, I.~D.~Moore, S.~Brudney, S.~Das, D.~T.~C.~Allcock and Y.~N.~Joglekar,
Observing super-quantum correlations across the exceptional point in a single, two-level trapped ion,
\href{https://arxiv.org/abs/2304.12413}
{arXiv:2304.12413}.

%===========================================================================
%===========================================================================
%===========================================================================
%===========================================================================
%===========================================================================
%===========================================================================
%===========================================================================
%===========================================================================


\end{thebibliography}
\end{document}